\documentclass[reprint,prd,nofootinbib]{revtex4-1}
\usepackage{graphicx}
\usepackage{xcolor}
\usepackage{amsmath}
\usepackage{siunitx}
\usepackage{subfigure}
\usepackage{multirow}
\usepackage[hidelinks]{hyperref}
\def\equationautorefname#1#2\null{%
  Eq.\;(#2\null)%
}
\def\figureautorefname#1\null{%
  Fig.#1\null
}

\begin{document}

\title{Impact of slow conversions on hybrid stars with sequential QCD phase transitions}
\author{Victor P. {\sc Gon\c{c}alves$^1$}}
\author{Lucas {\sc Lazzari$^1$}}

\affiliation{High and Medium Energy Group, Instituto de F{\'i}sica e Matem{\'a}tica,  Universidade Federal de Pelotas \\
  Caixa Postal 354,  96010-900, Pelotas, RS, Brazil}

\begin{abstract}
In this paper, we investigate the impact of the phase conversion speed on the properties of hybrid stars with sequential sharp QCD phase transitions. We consider that these hybrid stars possess an inner (outer) core of CFL (2SC) quark matter surrounded by hadronic matter. Assuming that the phase conversions can be slow or rapid, we analyze the dynamical stability of these objects. In particular, we present our predictions for the mass-radius profile and the fundamental eigenfrequencies of hybrid stars with sequential QCD first order phase transitions. Our results demonstrate that the usual stability criteria i.e. $\partial M/ \partial P_0 \geq 0$, is sufficient only if both conversions are rapid. Moreover, we show that if any of the interfaces has a slow conversion speed, the fundamental eigenfrequencies are significantly modified.
\end{abstract}



\maketitle

\section{Introduction}
\label{sec:intro}

One of the main challenges of the strong interaction theory -- Quantum Chromodynamics (QCD) -- is the description of matter at high temperatures and densities \cite{Pasechnik:2016wkt}. Under these extreme conditions, QCD predicts the  transition of hadronic to new partonic phases. This gives rise to the possibility that neutron stars (NSs) possess a quark matter core surrounded by a nuclear envelope (the so-called hybrid stars)~\cite{Alford:2004pf,alford2013,annala2020}. 

The subject of hybrid stars has gained more attention since the detection of the gravitational waves from the coalescence of two NSs in the GW170817 event~\cite{ligo2017}, which puts constraints on the equation of state (EOS) and radii~\cite{abbott2018}, as well as on the maximum mass~\cite{Margalit:2017dij, Rezzolla:2017aly} and tidal deformability~\cite{Raithel:2018ncd} of NSs. Many authors interpreted this remarkable event as an evidence for the presence of quark matter in the core of such stars~(see \cite{Bauswein:2018bma,annala2020} and references therein). Furthermore, the discovery of the most massive pulsar, PSR J0740+6620~\cite{NANOGrav:2019jur, Fonseca:2021wxt} put additional constraints on the NS EOS, which has to be able to produce such massive stars to be considered valid. Furthermore, the recent results from the NICER mission, which measured the radii of some NSs to unprecedented precision, added further constraints to the mass-radius relation of these objects~\cite{Miller:2019cac, riley2019, Miller:2021qha, Riley:2021pdl}. 

When one considers a sharp interface (Maxwell construction) between hadronic and deconfined quark matter inside the hybrid star, one has to account for effects such as the phase conversion speed between phases. In Refs.~\cite{haensel1989, Pereira:2017rmp}, the extra boundary conditions for sharp phase transitions were derived, based on the comparison between the timescales of the perturbations from radial oscillations and the phase conversion. These results were applied in Ref.~\cite{Lugones:2021bkm} considering only one phase transition, indicating that hybrid objects in the unstable branch become stable if the conversion speed is slow. Such objects, which they called `slow stable hybrid stars', satisfies all current and aforementioned astrophysical constraints. 

In the high density regime, present inside a NS, quark matter is expected to be in a color superconductor state (see Ref.~\cite{Alford:2007xm} for a review), which is believed to present two distinct phases: the two-flavor color superconductor (2SC) one, where one has the pairing of  up and down quarks; and the color-flavor-locked (CFL) one where  the pairing can occur between quarks of all colors and light flavors. An interesting possibility is  that sequential QCD phase transitions occur inside a NS~\cite{Alford:2017qgh,Han:2018mtj,Rodriguez:2020fhf}. In this sense, the first transition would be from hadronic to 2SC, and the second from 2SC to CFL quark matter. This implies that the EOS used to describe quark matter has two distinct sound speeds, being the CFL phase EOS stiffer than the 2SC one.  Regarding the quark matter EOSs, all aforementioned references used the constant speed of sound (CSS) parametrization~\cite{Alford:2017qgh}, assuming that both phase transitions are sharp. In Refs.~\cite{Alford:2017qgh, Han:2018mtj}, the usual stability criteria was used to determine stable configurations in the sequential QCD phase transitions scenario. In Ref.~\cite{Rodriguez:2020fhf}, the authors have studied the $g_2$ mode within the Cowling approximation and the dynamical stability of these objects\footnote{Both Refs.~\cite{Han:2018mtj,Rodriguez:2020fhf} and Ref.~\cite{Li:2019fqe} have considered the impact of the second phase transition on tidal deformability.}. However, a single conversion speed was assumed in these previous studies. 


In this paper, we will investigate the impact of the phase conversion speed on the properties of hybrid stars with sequential QCD phase transitions and analyze the dynamical stability of these stars considering different combinations for the phase conversion speed. Our goal is to estimate the impacts of slow phase conversions and the stable hybrid configurations that arise in these different scenarios.

This paper is organized as follows. In the next section, we will present a brief review of the formalism used to describe hybrid stars with sequential QCD phase transitions and to estimate the dynamical stability assuming slow and rapid conversions. In~\autoref{sec:results}, we present our results for the mass-radius profile and for the fundamental mode of oscillation considering different combinations of conversion speeds for the  two sequential phase transitions. Finally, in~\autoref{sec:sum} we summarize our main conclusions. In this paper, we use geometric units where $c = G = 1$.

\section{Formalism}
\label{sec:theory}
In what follows, we will assume that quark matter appears in two distinct phases inside hybrid stars: a 2SC phase (outer core) and a CFL one (inner core). In this sense, there is two QCD phase transitions taking place inside the hybrid star, one from nuclear matter to deconfined quark matter in a 2SC phase, and a subsequent one from 2SC to a CFL phase. We will assume both phase transitions to be of first order, i.e. that the pressure is constant but there is a jump in the energy density. 

The EOS for nuclear matter will be modelled using generalized piece-wise polytropes (GPP) following the prescription of Ref.~\cite{OBoyle:2020qvf}, which assures continuity in all thermodynamic quantities. In particular, we use the Sly4 crust, as presented in Table II of Ref.~\cite{OBoyle:2020qvf}. Inspired by Ref.~\cite{Lugones:2021bkm}, we use  model-agnostic GPPs for densities larger than 0.3 $n_0$, such that it matches  the upper limit of the  EOS calculated from chiral effective field theory in Ref.~\cite{Hebeler:2013nza} at 1.1 $n_0$,  where $n_0 = \SI{0.16}{fm^{-3}}$ is the nuclear saturation density. Distinctly from Ref.~\cite{Lugones:2021bkm}, we assume that the hybrid stars share the same hadronic branch, with the parameters that describe the GPPs being given in Table~\ref{tab:eoss}.

\begin{figure}[!t]
  \centering
  \includegraphics[width=0.525\textwidth]{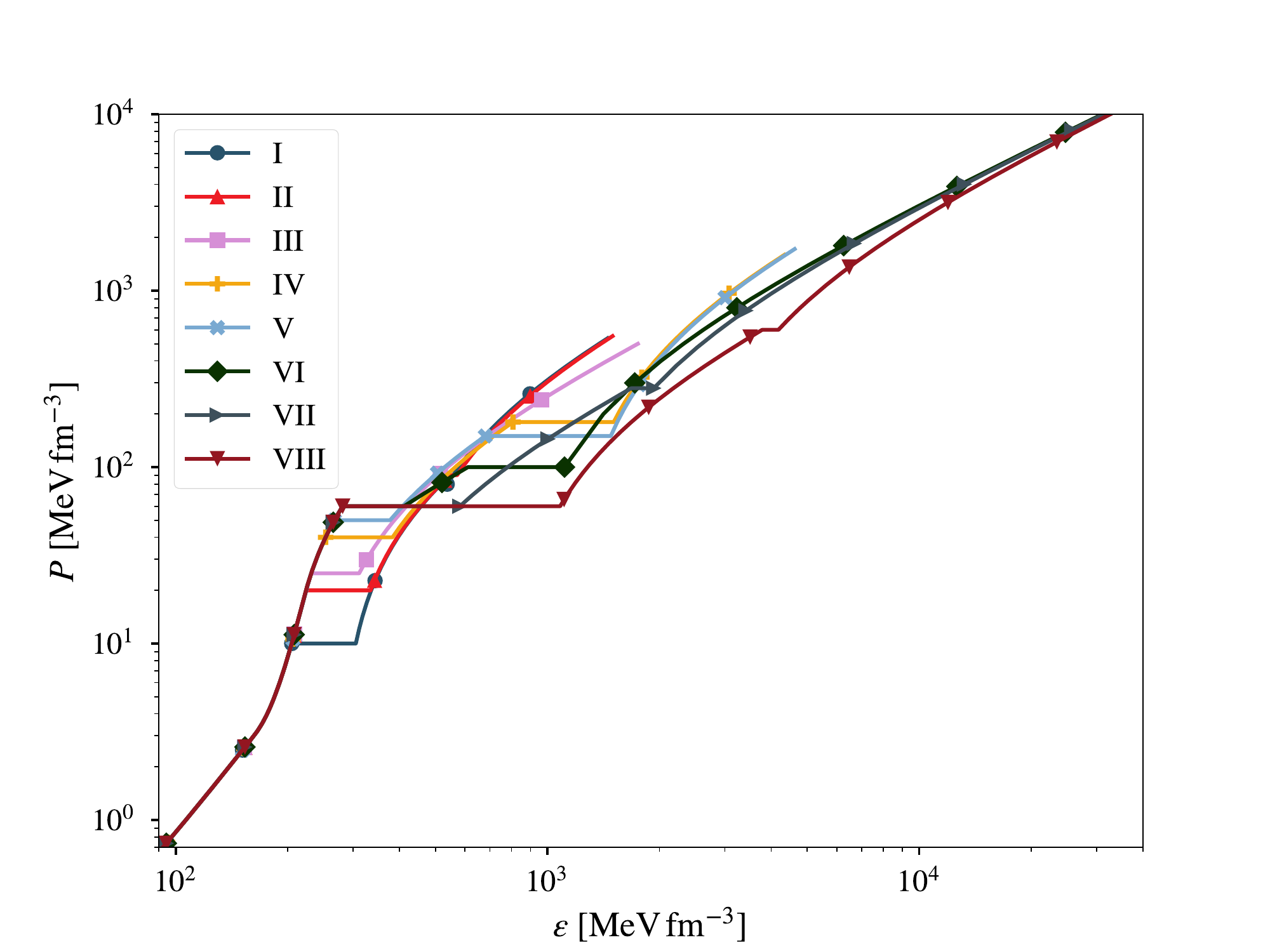}
  \caption{\label{fig:eoss} Hybrid EOSs with two sequential QCD first order phase transitions considered in our analysis.}
\end{figure}

\begin{figure}[!t]
    \centering
    \includegraphics[width=0.525\textwidth]{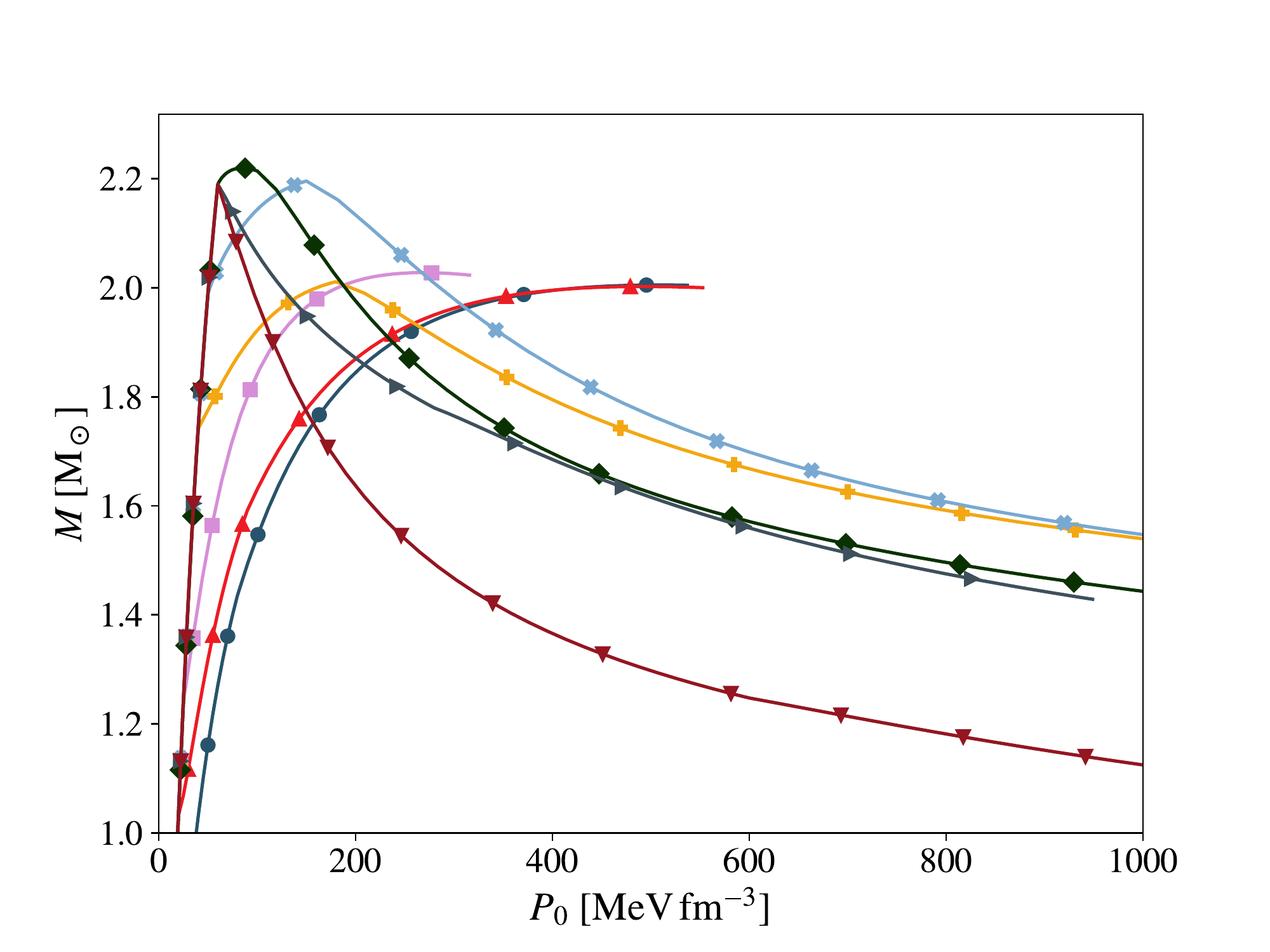}
    \caption{\label{fig:MxP0} Mass as a function of the central pressure for the EOSs presented in Fig.~\ref{fig:eoss}}
\end{figure}

On the other hand, the quark matter EOS with sequential QCD phase transitions will be described using a generalization of the CSS parametrization, where the speed of sound is assumed constant in each phase, with the resulting EOSs being linear. Following  Ref.~\cite{Alford:2017qgh}, the QM EOSs will be given by
\begin{equation}
  \label{eq:QM_EOS}
  P(\epsilon) =
  \begin{cases}
    P_1\,, \hspace{\fill} \epsilon_1 < \epsilon < \epsilon_1 + \Delta \epsilon_1\,,\\
    P_1 + s_1[\epsilon - (\epsilon_1 + \Delta \epsilon_1)]\,, \hspace{\fill} \epsilon_1 + \Delta \epsilon_1 < \epsilon < \epsilon_2\,, \\
    P_2\,, \hspace{\fill} \epsilon_2 < \epsilon < \epsilon_2 + \Delta \epsilon_2\,,\\
    P_2 + s_2[\epsilon - (\epsilon_2 + \Delta \epsilon_2)]\,, \hspace{\fill}  \epsilon > \epsilon_2 + \Delta \epsilon_2\,.
  \end{cases}
\end{equation}
One has that this parametrization describes the sequential phase transitions in terms of six parameters. We have chosen as free parameters the transitional pressures $P_i$, jump in energy density $\Delta \epsilon_i$ and the sound speed squared $s_i$, where $i = 1, 2$ represents the first (2SC) and second (CFL) quark phases, respectively. The results obtained in Refs.~\cite{alford2013,Alford:2015dpa} indicate that $P_1$ determines the maximum mass of the hadronic branch, while $P_2$ determines the maximum mass of the first hybrid branch. The jump in energy density defines the maximum mass of the respective hybrid configurations. Hereafter, we will assume that the first (second) phase transition corresponds to the nuclear - 2SC (2SC - CFL) transition. 
Note that causality requires $s_i < 1$. As the CFL phase is expected to be characterized by a stiffer EOS than the 2SC phase~\cite{Alford:2017qgh}, we will also assume that $s_2 > s_1$. In order to estimate the impact of the speed of sound in our analysis, we chose two conservative sets of values. The first one is based on phenomenological models, where the quark matter EOS speed of sound squared tends to the conformal limit (1/3). For this, we set $s_1 = 0.2$ and $s_2 = 0.33$. For the second set, we explore the possibility that quark matter is stiffer than suggested by current models, and assume that $s_1 = 0.33$ and $s_2 = 0.5$. The remaining parameters where evaluated over the following ranges (all in \si{MeV\,fm^{-3}}):
\begin{align*}
10 \leq P_1 \leq 60\,, & \qquad 80 \leq P_2 \leq 600\,, \\
80 \leq \Delta \epsilon_1 \leq 800 \,, & \qquad 20 \leq \Delta \epsilon_2 \leq 800\,.
\end{align*}
This way, we have computed more than 4000 EOSs, keeping only the ones that produce stellar configurations that surpass the two solar mass limit. From all valid EOSs, we chose 8 representative ones which are presented in Fig.~\ref{fig:eoss}, with the associated parameters given in Table \ref{tab:eoss}. The distinct models are characterized by different values of the transition pressures. In Fig.~\ref{fig:MxP0}, we present the gravitational mass as a function of the central pressure, where one can analyze  the regions that  satisfy the usual stability criteria $\partial M/ \partial P_0 \geq 0$. In particular, models I, II and III correspond to a small  ($\leq \SI{25}{MeV\,fm^{-3}}$) first transition pressure. Unlike the first two, model III presents a high value of $P_2$ and $\Delta \epsilon_2$. In contrast, from model IV to VIII, the first transition pressure is high. As shown in Fig.~\ref{fig:MxP0}, for models VI, VII and VIII one has that the hadronic branch already reaches the 2 solar mass limit, which allows  quark matter to be characterized by smaller values of the speed of sound. 

\begin{table}[!t]
  \begin{center}
    \begin{tabular}{ccccccc}
    \hline
    \multicolumn{7}{c}{{\bf Hadronic EOS}}
    \\ \hline
     $\log_{10} K_1$ & $\Gamma_1$ & $\Gamma_2$ & $\Gamma_3$ & $\log_{10} \rho_1$ & $\log_{10} \rho_2$ & \\
    \hline
    -27.22 & 2.764 & 10.0 & 2.0 & 14.45 & 14.58 &\\
    \hline
    \hline
    \multicolumn{7}{c}{{\bf Quark Matter EOSs}} \\
    \hline
      EOS &  $P_1$ &  $\Delta \epsilon_1$ & $s_1$ & $P_2$ &  $\Delta \epsilon_2$ & $s_2$
      \\
      \hline
      \hline
      I & 10 & 100 & \multirow{5}{*}{0.33} & 80 & 20 & \multirow{5}{*}{0.50} \\
      \cline{1-3}\cline{5-6}
      II & 20 & 110 &  & 90 & 25 &  \\
      \cline{1-3}\cline{5-6}
      III & 25 & 80 &  & 500 & 600 &  \\
      \cline{1-3}\cline{5-6}
      IV & 40 & 130 &  & 180 & 700 & \\
      \cline{1-3}\cline{5-6}
      V & 50 & 110 &  & 150 & 800 &  \\
      \hline
      VI & \multirow{3}{*}{60} & 130 & \multirow{3}{*}{0.2} & 100 & 500 & \multirow{3}{*}{0.33} \\
      \cline{1-1}\cline{3-3}\cline{5-6}
      VII &  & 300 & & 280 & 250 &  \\
      \cline{1-1}\cline{3-3}\cline{5-6}
      VIII & & 800 &  & 600 & 400 &  \\
    \end{tabular}
  \end{center}
  \caption{Transition pressure and jump in the energy density (both in~\si{MeV\,fm^{-3}}) for both phase transitions and respective quark matter speed of sound squared for the 2SC and CFL phases.}
  \label{tab:eoss}
\end{table}

In order to investigate the dynamical stability of  hybrid stars with sequential phase transitions, one has to solve the pulsation equation derived by Chandrasekhar~\cite{chandra1964}, which determines the stable configurations as being those that have a {positive squared fundamental eigenfrequency $\omega^2$ and unstable if $\omega^2 < 0$}. For numerical purposes it is preferable to use the Gondek-Rosinka's form~\cite{Gondek:1997fd}, that implies that the pulsation equation can be expressed in terms of two first-order equations for the relative radial displacement $\xi \equiv \Delta r/r$ and the Lagrangian perturbation of the pressure $\Delta P$. These coupled differential equations are given by
\begin{align}
  \label{eq:dxidr}
  \frac{d\xi}{dr} & = -\frac{1}{r}\left(3\xi + \frac{1}{\Gamma P}\right) + \frac{d\nu}{dr}\xi \,,\\
  \frac{d\Delta P}{dr} & = \xi\left[e^{2\lambda}(\omega^2e^{-2\nu} - 8\pi P) + \frac{d\nu}{dr}\left(\frac{4}{r}+\frac{d\nu}{dr}\right)\right](P+\epsilon)r \nonumber \\
                  & \quad -\Delta P\left[\frac{d\nu}{dr} + 4\pi(P+\epsilon)r e^{2\lambda}\right]\,,
\end{align}
where $\Gamma = (1 + \epsilon/P)s$ is the adiabatic index and the quantities $\xi$ and $\Delta P$ are assumed to have a harmonic time dependence. The  coefficients of this system of equations are obtained by solving the Tolman-Oppenheimer-Volkoff (TOV) equations for a given central pressure. It is also necessary to stipulate boundary conditions such as $(\Delta P)_{r=0} = -3(\xi\Gamma P)_{r=0}$ on the stellar center and $(\Delta P)_{r=R} = 0$ on the surface. We also consider normalized eigenfunctions $\xi(0) = 1$.  As we are considering that two sharp phase transitions are present, we also must to assume junction conditions to describe the interface of different phases. As discussed in detail in Ref.~\cite{Pereira:2017rmp},  these junction conditions are related to the velocity of the phase transition near the surface splitting the two phases. 
If the time needed to convert one phase is smaller than those of the perturbations, the nature of volume elements from one phase into another changes instantaneously near the sharp interface and we have a rapid phase transition \footnote{ The oscillation period of the fluid elements close to the quark-hadron interface is determined by the fundamental frequency $\omega_0$, which is of the order of $\approx 1 - 100$ kHz. Therefore, rapid and slow conversions are characterized by a timescale smaller or larger than $\approx 0.01 - 1$ ms (See e.g. \cite{haensel1989,Lugones:2021zsg}).}. In this case, the splitting surface can be assumed to be in thermodynamic equilibrium, which results in $\Delta P^+ - \Delta P^- = 0$, where the + (-) represents the Lagrangian perturbation of the pressure after (before) the phase transition. However, the relative radial displacement is no longer continuous due to the fast conversion of elements near the surface. In Ref.~\cite{Pereira:2017rmp}, it was shown that the appropriate extra boundary condition for $\xi$ is
$$\xi^+ = \xi^- + \frac{\Delta P}{r}\left(\frac{1}{P_0^{'+}} - \frac{1}{P^{'-}_{0}}\right) \,,$$ where the prime denotes the radial derivative and $P_0 \equiv P_0(r)$ is the unperturbed pressure. In contrast, in slow phase transitions, one has that the characteristic timescale of the transition of one phase into another is much larger than those of the perturbation. In this case, volume elements near the sharp interface do not change their nature due to the oscillations, but rather co-move with it. As a consequence, the jump in $\xi$ across the splitting surface is null. As demonstrated in Ref.~\cite{Pereira:2017rmp}, the junction condition $\Delta P^+ - \Delta P^- = 0$ is also valid for slow phase transitions. 

The results derived in Ref.~\cite{Pereira:2017rmp} indicated that the dynamical stability of hybrid stars is strongly influenced by the speed of the phase conversion. For slow conversions the frequency of the fundamental mode can be a real number even for branches of stellar models that verify $\partial M/\partial P_0 < 0$, i.e. stable stellar configurations are possible even when the usual stability condition is not satisfied. In contrast, for rapid phase conversions one has that 
real frequencies only occur for  $\partial M/\partial P_0 \geq 0$. As a consequence, slow phase conversions imply the existence of a new branch of stable stellar configurations. In Ref. \cite{Lugones:2021bkm}, the authors demonstrated that the stars on the $\partial M/\partial P_0 < 0$ branch  satisfy all  current observational constraints. In the next Section, we will analyze the dynamical stability of hybrid stars with sequential QCD phase transitions by solving the TOV and pulsation equations assuming different combinations of conversion speeds.

\section{Results}
\label{sec:results}

In our analysis, we will consider hybrid stars with two sequential first order phase transitions, which can be slow or rapid. Previous studies have considered that both sharp transitions were characterized by the same conversion speed~\citep{Alford:2017qgh,Han:2018mtj,Rodriguez:2020fhf}. In our work, we will also present the configurations that can be analysed through the usual stability criteria, which considers that both conversions are rapid and will be denoted by rapid-rapid (rr) hereafter. Slow conversions can occur in one or both phase transitions. When a slow conversion is present in only one of the phase transitions inside the star, we will consider the  slow-rapid (sr) and rapid-slow (rs) combinations, which denote the speed of conversion for the transition from nuclear matter to quark matter and from 2SC to CFL, respectively. In this sense, we consider a sequence of transitions from the surface to the core of the star.  Finally, we will also investigate the case where both phase transitions are slow, and the predictions will be denoted by slow-slow (ss). As pointed out in the previous Section, in the slow scenario, dynamical stability does not coincide with the classical condition $\partial M/\partial P_0 \geq 0$, in contrast to rapid ones. Considering slow transitions, the configurations that are stable but present a negative mass derivative will be called slow-stable. If they are both rapid- and slow-stable they will be called totally stable.  

\begin{figure*}[!t]
  \centering
  \subfigure{\includegraphics[width=0.32\textwidth]{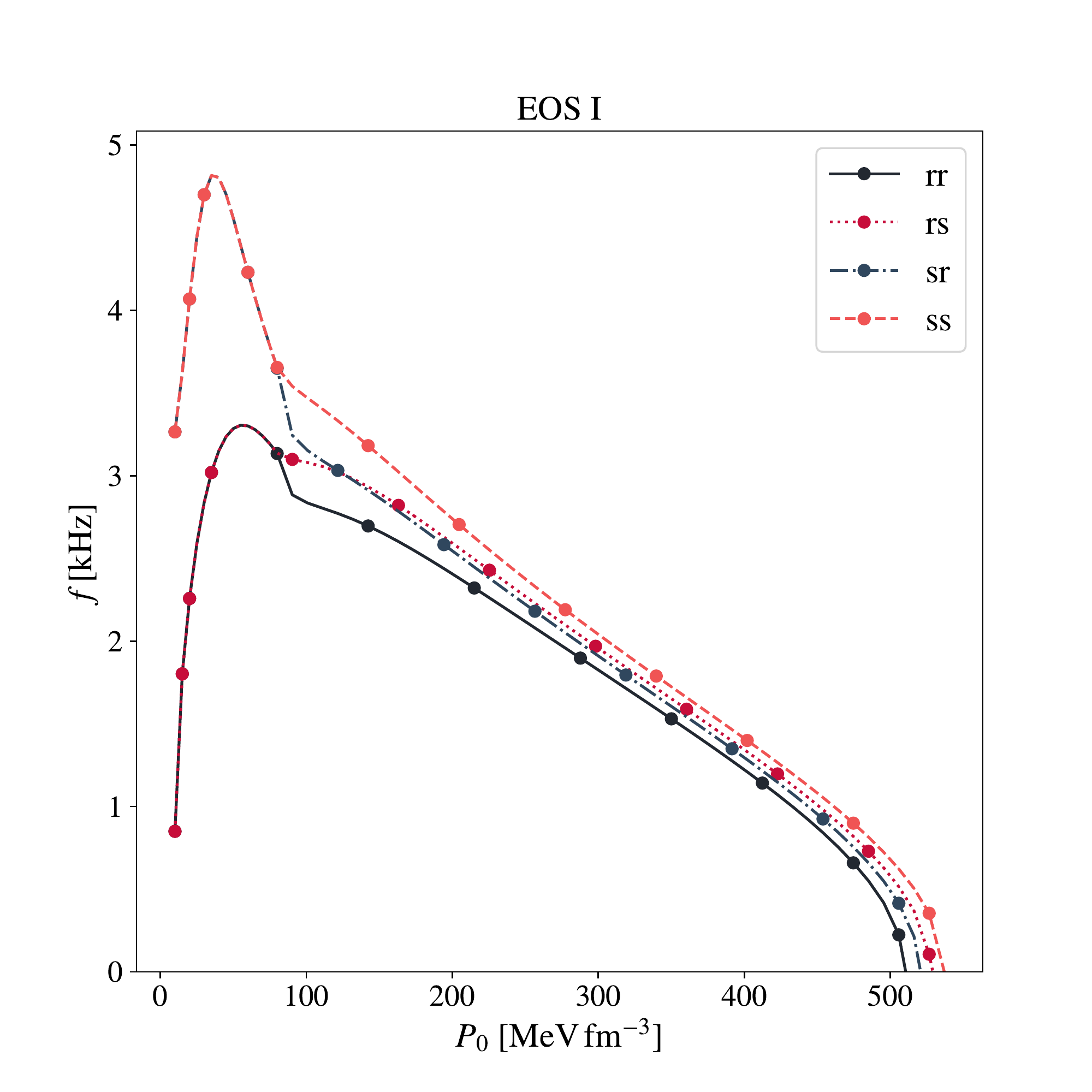}}
  \subfigure{\includegraphics[width=0.32\textwidth]{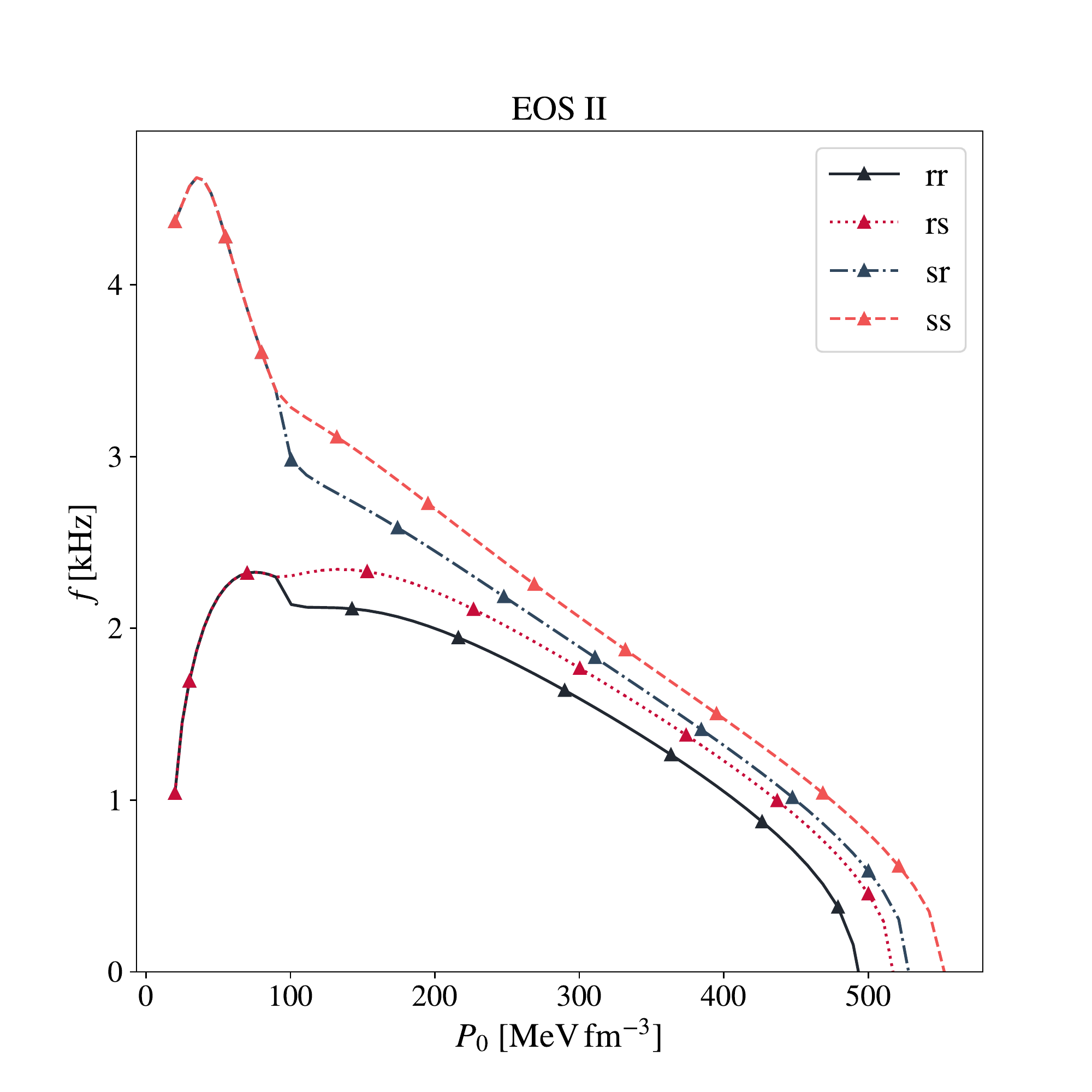}}
  \subfigure{\includegraphics[width=0.32\textwidth]{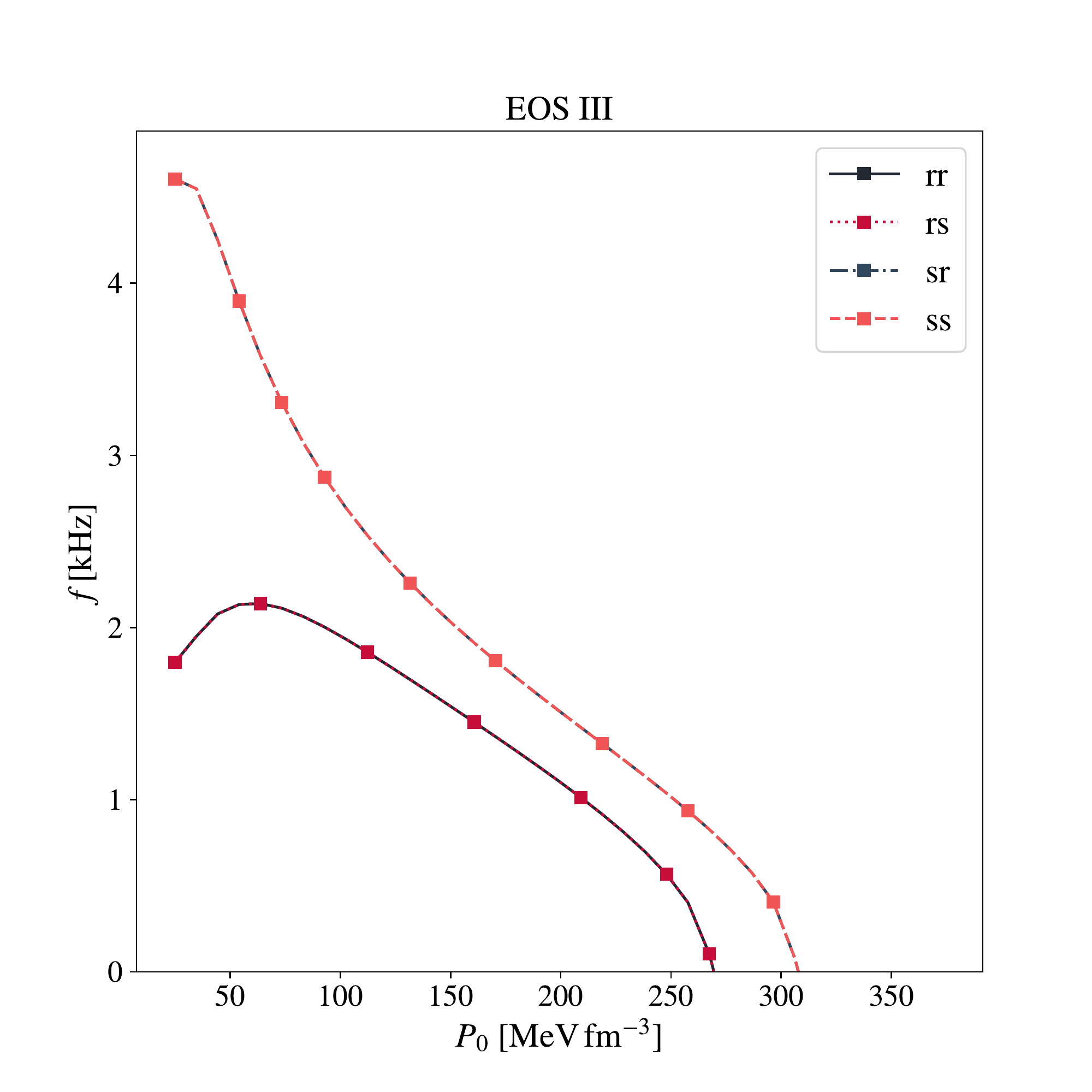}}
  \subfigure{\includegraphics[width=0.32\textwidth]{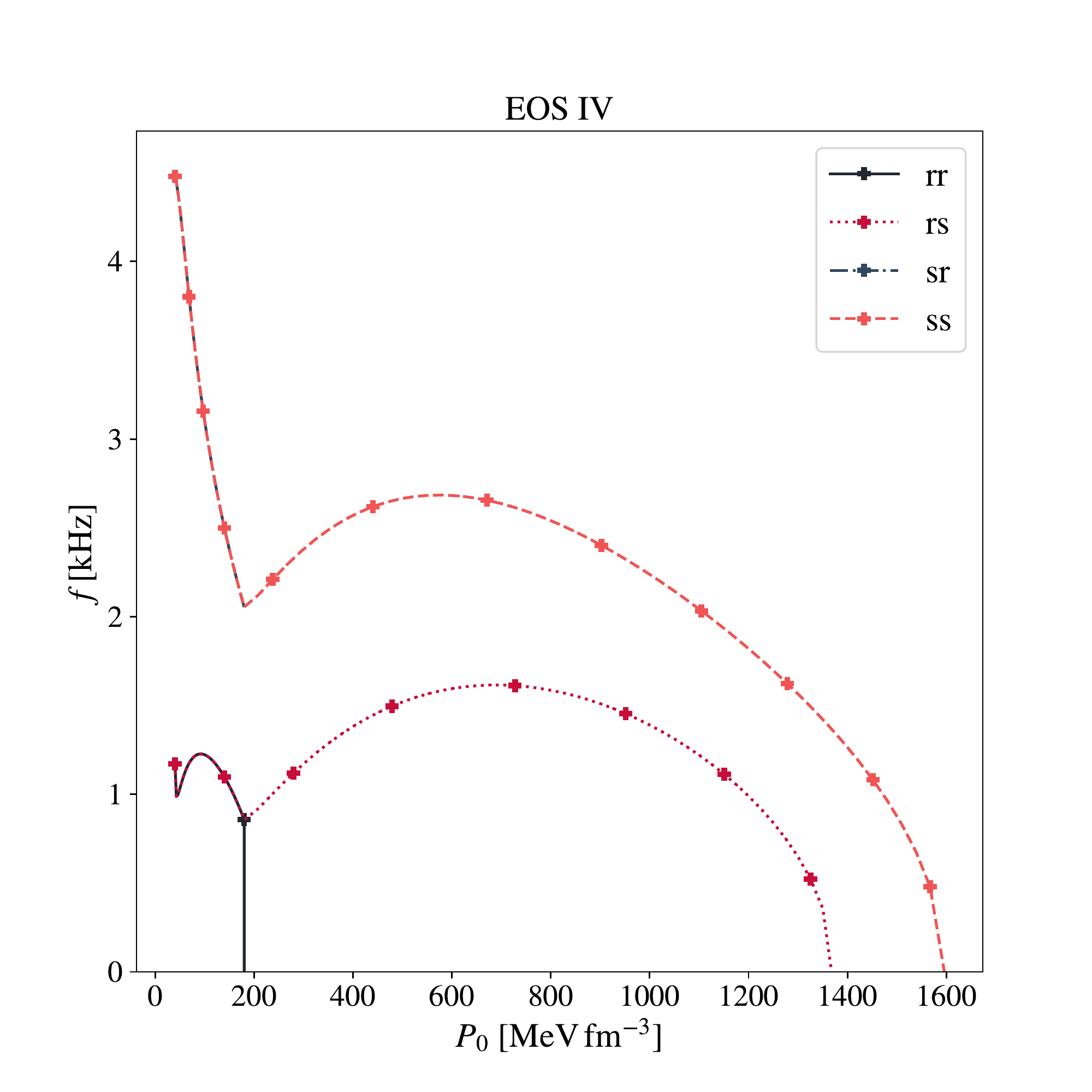}}
  \subfigure{\includegraphics[width=0.32\textwidth]{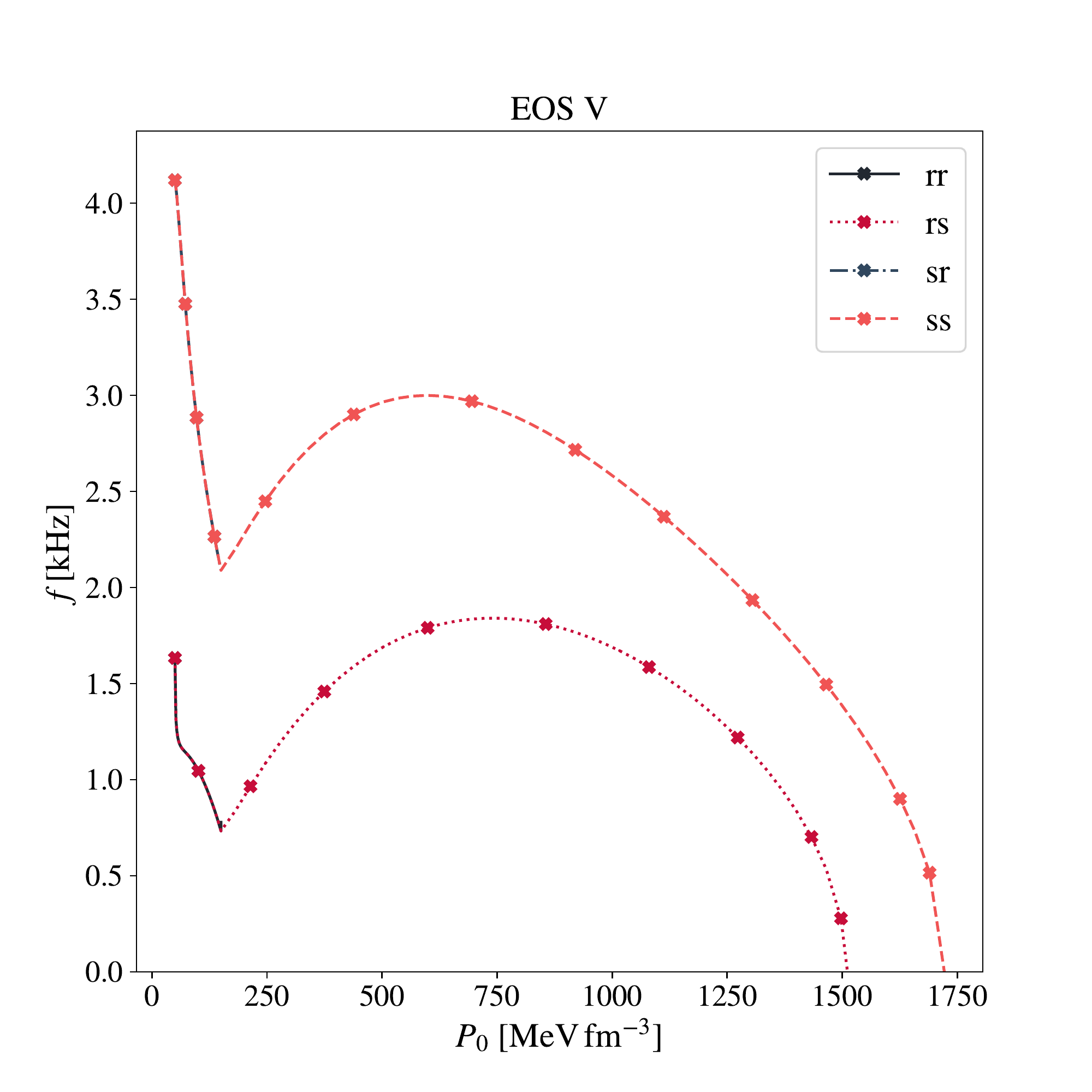}}
  \subfigure{\includegraphics[width=0.32\textwidth]{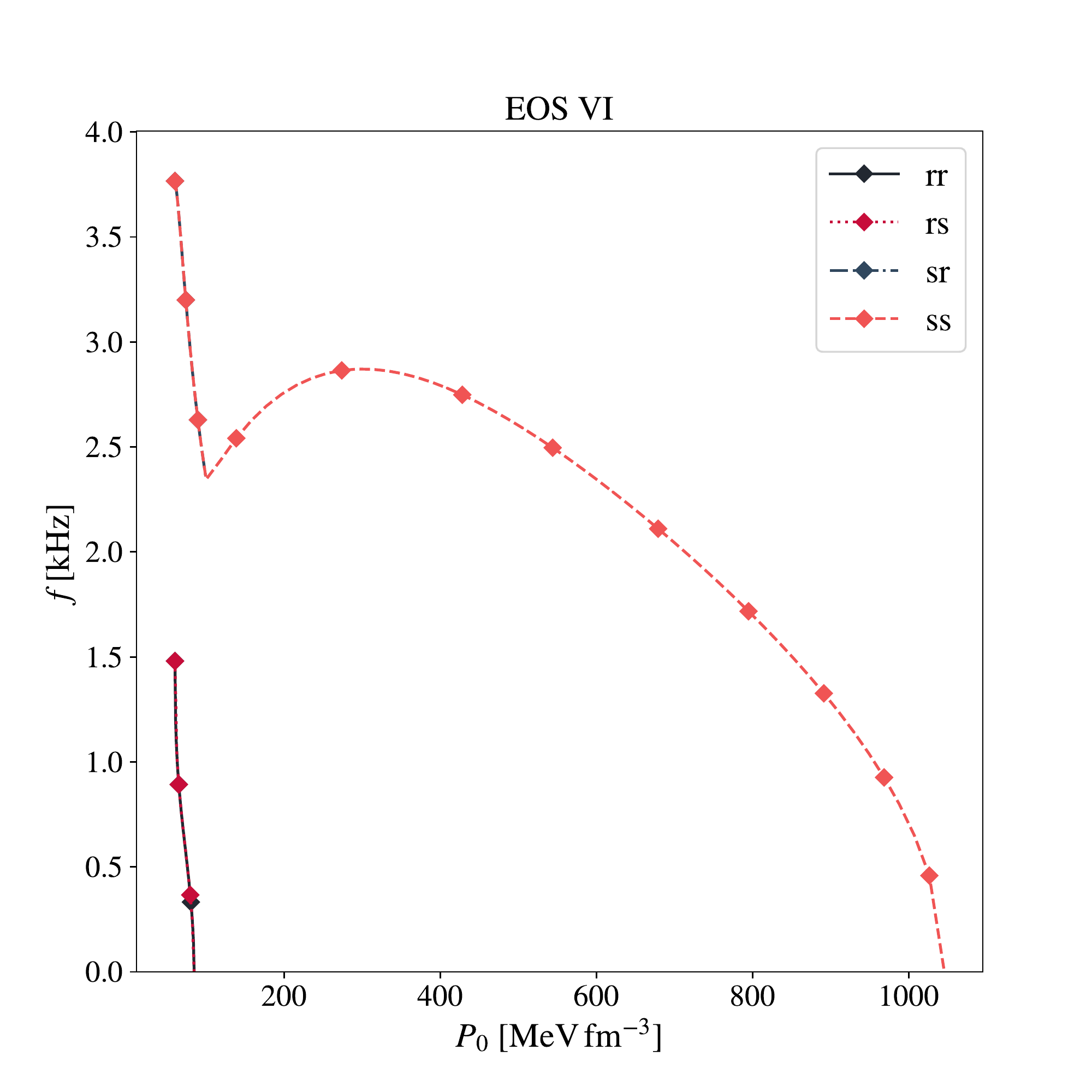}}
  \subfigure{\includegraphics[width=0.32\textwidth]{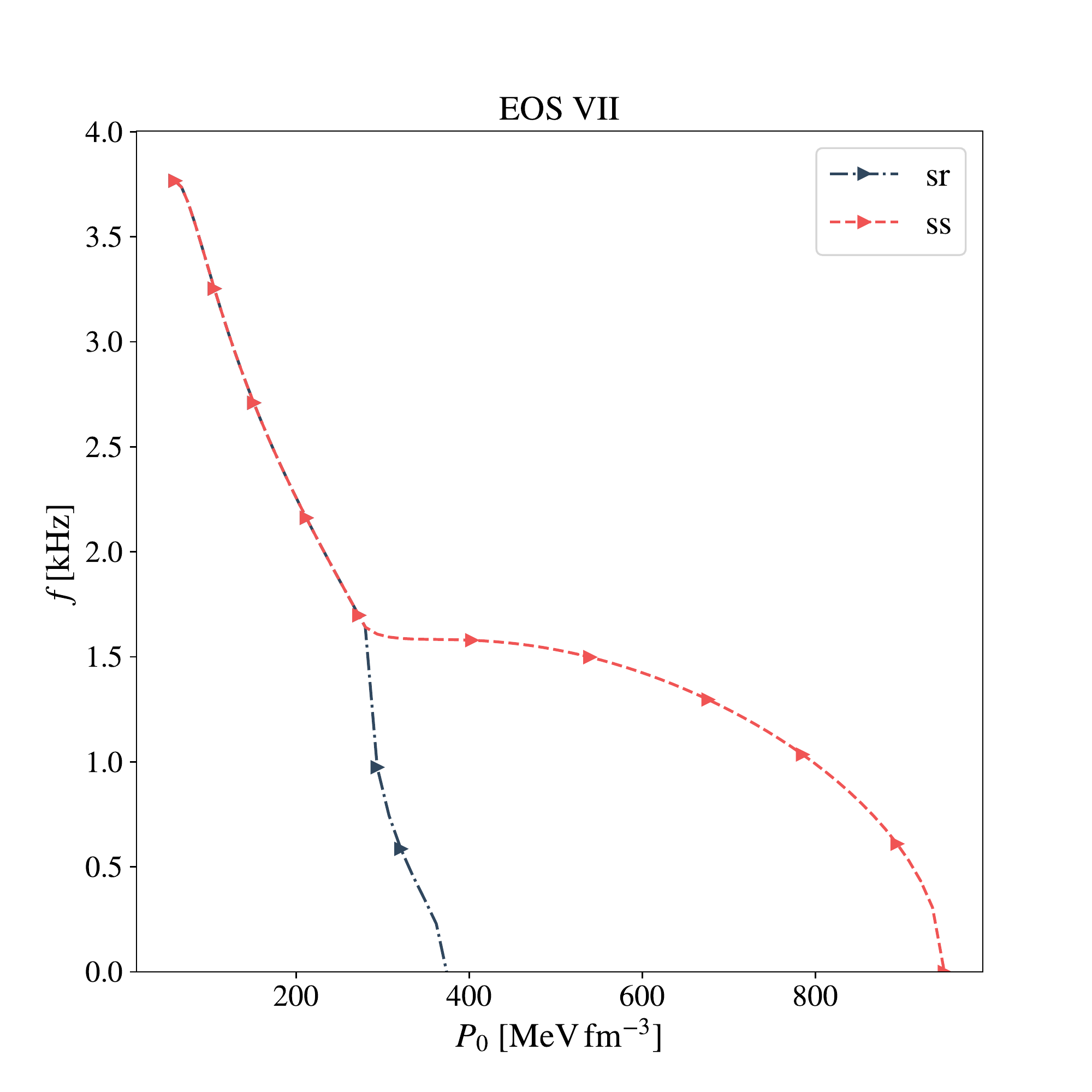}}
  \subfigure{\includegraphics[width=0.32\textwidth]{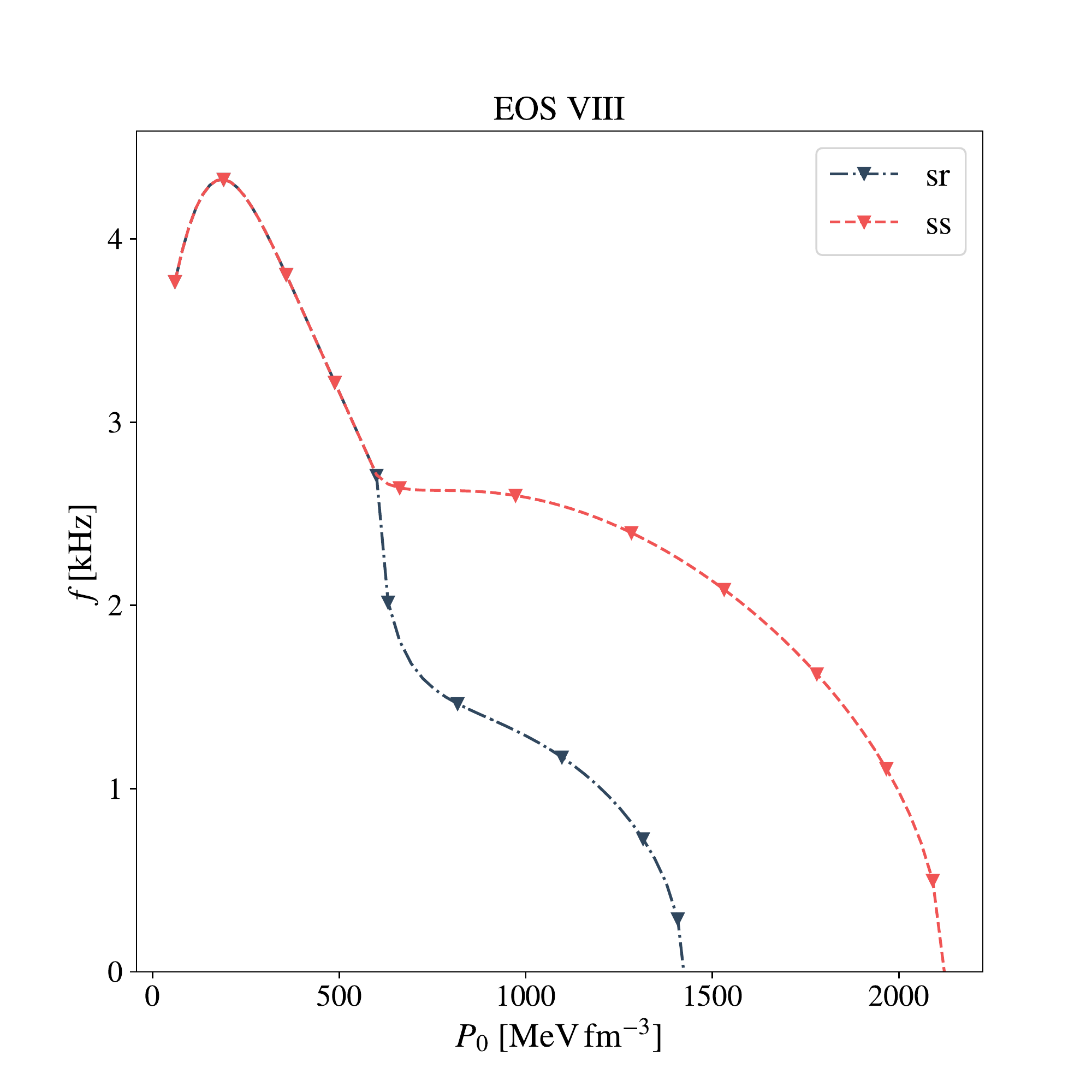}}
  \caption{\label{fig:fxP0} Fundamental linear eigenfrequency as a function of central pressure, for all models considered and for the different combinations of transition speeds. }
\end{figure*}

Initially, lets investigate the fundamental eigenfrequency of HSs for some models considering different combinations of transition speeds. In Fig.~\ref{fig:fxP0}, we present the fundamental eigenfrequency as a function of the central pressure for the selected EOS. The labels `r' and `s' stand for rapid and slow transitions, respectively. 
The results for EOSs I and II present a clear similarity, since in both cases the first transition happens for small pressures, and the difference between $P_2$ and $P_1$ is not very large. One has that  both models present many configurations that are totally stable. However, when one of the conversion speeds is slow we have that configurations beyond the maximum mass are slow-stable. This happens because the Lagragian perturbations are altered and lead to higher frequencies in the slow scenario, which ends up pushing the last stable configuration to larger central pressures even if one of the conversions is rapid. For model III, we have found that the stable configurations do not reach high enough central pressures for the occurrence of the second phase transition, even in the `ss' case. 
For models IV and V, the difference between the transition pressures is not very large, however $\Delta \epsilon_2$ is considerably large. Consequently, the second hybrid branch has a negative mass derivative, as can be seen in Fig.~\ref{fig:MxP0}. This implies that only the first hybrid branch is totally stable. However, if the second phase transition has a slow conversion, many slow-stable configurations occur beyond the maximum mass. These models suggest that if the second phase transition is slow, many configurations with very high central pressures (densities) are slow-stable. 

The impact of the speed of sound can be investigated by analyzing the predictions  of the Models VI -- VIII, which assume  $s_1 = 0.2$ and $s_2 = 0.33$ . Model VI is similar to models IV and V and has a first hybrid branch that is totally stable against radial oscillations. In contrast to models IV and V, model VI does not present slow-stable configurations in the 'rs' case. Regarding models VII and VIII, which have high transition pressures and energy density jumps, we clearly see that they do not present any totally stable configurations. This was expected from Fig.~\ref{fig:MxP0}, where the hybrid configurations in these models do not satisfy the usual stability criteria, which also corresponds to the dynamical stability of rapid phase conversions.

\begin{figure*}[!t]
  \centering
  \subfigure{\includegraphics[width=0.45\textwidth]{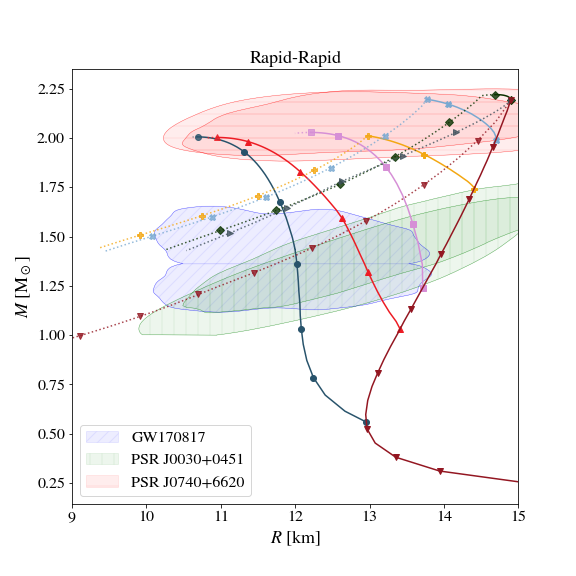}}
  \subfigure{\includegraphics[width=0.45\textwidth]{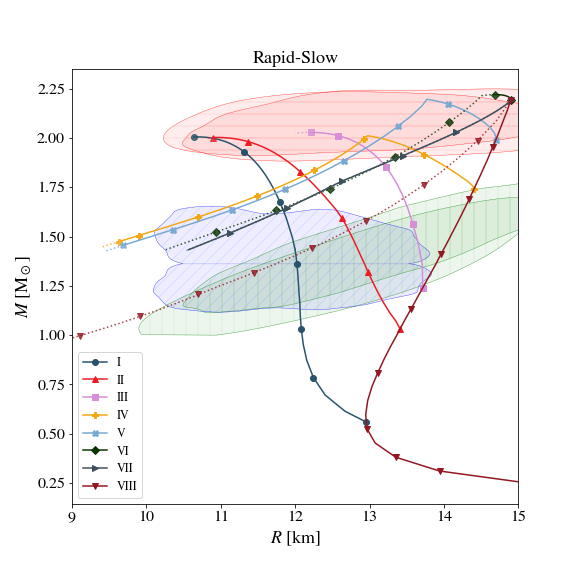}}
  \subfigure{\includegraphics[width=0.45\textwidth]{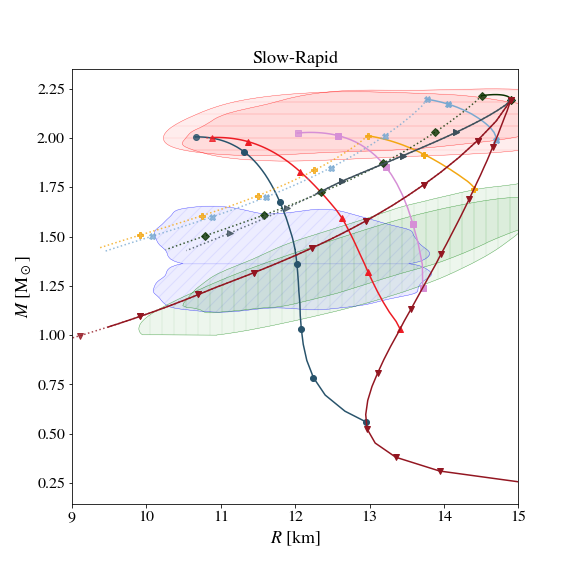}}
  \subfigure{\includegraphics[width=0.45\textwidth]{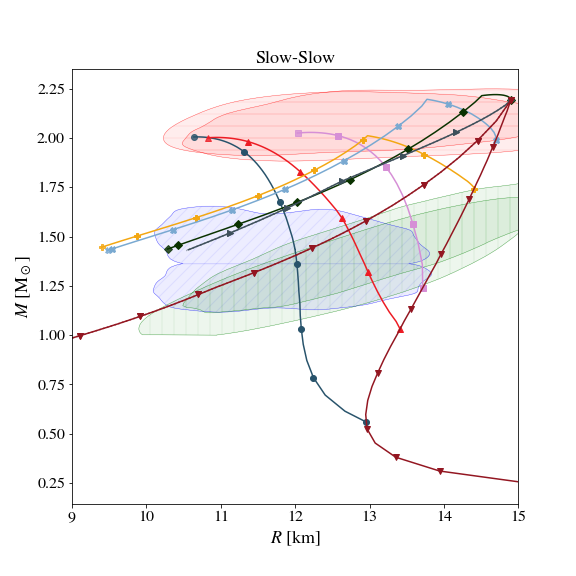}}
  \caption{\label{fig:MxR} Mass-radius profile of hybrid stars with sequential QCD first order phase transitions, considering different conversion rates. {The dotted lines represent the configurations that have $\omega_0^2 < 0$ in the respective conversion speed analysis. }}
\end{figure*}

The predictions for the mass-radius profile of a hybrid star with two sequential phase transitions are presented in Fig.~\ref{fig:MxR} considering the different combinations for the conversion speed of the transitions. The solid marked lines indicate stable configurations, i.e. configurations that have a real fundamental eigenfrequency, as presented in Fig.~\ref{fig:fxP0}. Initially, let's consider the rapid-rapid case presented in the top left panel of Fig.~\ref{fig:MxR}. As expected, in this case, the configurations that are rapid stable satisfy $\partial M/ \partial P_0 \geq 0$ and are called totally stable. All models up to model VI present such configurations. As one can clearly see, increasing $P_1$ leads to smaller stability regions in this rapid-rapid scenario. 

In the top right panel, we present our predictions considering that the transition from nuclear matter to quark matter is rapid, but the transition from 2SC to CFL is slow. This result is very similar to the previous case, where both phase transitions were rapid. The main difference is that a second hybrid branch appears in models IV and V. Since the second phase transition is slow, this new hybrid branch presents $\partial M/\partial P_0 < 0$ and is called a slow-stable one.  

Considering that the phase transition from hadronic matter to 2SC is slow, but the transition from 2SC to CFL is rapid we obtain the results presented in the left bottom panel. In this case, all models present stable configurations, even the ones that didn't appear before (models VII and VIII). At the same time, the second hybrid branch of model IV and V is no longer stable, since the second transition is now a rapid one. 

Finally, in the right bottom panel, we assume that both transitions are slow. In this case, many regions where $\partial M/\partial P_0 < 0$ are slow-stable. In fact, all hybrid configurations for EOSs VII and VIII have a negative mass derivative but are slow-stable in both phase transitions. In particular, configurations that satisfy all current mass-radius constraints are present. Furthermore, the configurations obtained from EOSs V, VI and VII are totally stable up to the second phase transition and become only slow-stable after the maximum mass. As was presented in Ref.~\cite{Lugones:2021bkm}, the slow scenario opens up the parameter space for quark matter, and many configurations previously thought not to exist are now possible. As we have shown, this also happens when two phase transitions occur inside the star.

\section{Summary}
\label{sec:sum}

In this paper, we have presented the main implications of sequential QCD first order phase transitions on the dynamical stability of compacts hybrid stars. We showed that for low values of the first phase transition pressure (from nuclear to 2SC quark matter) there are many absolutely stable models that satisfy current mass-radius constraints. In this case, the phase transition from 2SC to CFL only leads to major changes in the star's properties if the jump in energy density is high enough. For small values of $\Delta\epsilon_2$ the hybrid configurations with one or two phase transitions are very similar, and practically indistinguishable from the mass-radius relation. However, an analysis of the $f$-modes of hybrid stars may provide substantial evidence for sequential phase transitions, which may be proved in future experiments. In this regard, future astrophysical data can provide useful information about the characteristic conversion rates from hadronic matter to quark matter and between its different phases. On the other hand, if the transition pressure from nuclear matter to quark matter is high, the only systems that can exist in Nature are the slow-stable ones. 

\section*{Acknowledgements}
The authors thank German Lugones and Jos\'e Jim\'enez for valuable discussions. This work was partially financed by the Brazilian funding agencies CNPq, Coordena\c{c}\~ao de Aperfei\c{c}oamento de Pessoal de N\'ivel Superior  (CAPES) -- Finance Code 001,   FAPERGS and INCT-FNA (process number 464898/2014-5).

\end{document}